\documentstyle{l-aa}
\begin{document}
\thesaurus{02.04.01, 08.14.1, 08.16.6}
\title{On the minimum period of uniformly rotating neutron stars}
\author{P. Haensel\inst{1,2}, J.P. Lasota\inst{1}\thanks{Present address: 
 Institut d'Astrophysique de Paris, CNRS, 98bis 
 Boulevard Arago, 75014 Paris, France}, and J.L. Zdunik\inst{2}}

\institute{Departement d'Astrophysique Relativiste et de Cosmologie, 
UPR 176 du CNRS, Observatoire de Paris, Section de Meudon,
 F-92195 Meudon Cedex, France
\and 
N. Copernicus Astronomical Center, Polish
           Academy of Sciences, Bartycka 18, PL-00-716 Warszawa, Poland\\
{\em e-mail: jlz@camk.edu.pl, haensel@camk.edu.pl, lasota@obspm.fr}}
\offprints{P. Haensel}
\date{}
\maketitle
\markboth{P. Haensel et al. 
                 Minimum period of uniformly rotating neutron stars}{} 
%==========================================================
\begin{abstract}
We show that the neutron star minimum period of uniform rotation is
determined, for causal equations of state, by the maximum value of the
relativistic (compactness) parameter $2GM/Rc^2$, allowed by causality
for static neutron stars, $0.7081$, and by the largest measured mass of
a   neutron star.  The relation between these three
quantities, resulting from the extrapolation  of  the empirical formulae
of Lasota et al. (1996), yields the minimum period of $0.288$~ms, only
2\% higher than an  absolute lower  bound obtained in extensive   exact
numerical calculations of Koranda et al. (1997).

\end{abstract}
%

%%%%%%%%%%%%%%%%%%%%%%%%%%%%%%%%%%
\section{Introduction }
The lower limit on the period of a uniformly rotating neutron star is
sensitive to the equation of state (EOS) of dense matter above the
nuclear density.  Therefore, an  uncertainty in the high density EOS
implies a large uncertainty in the minimum period of uniform rotation,
$P_{\rm min}$ (see, e.g. Friedman \& Ipser 1987; Friedman, Parker \&
Ipser 1989; Salgado et al. 1994a,b; Cook et al. 1994).  Hence, it is,
therefore, of interest to find a lower limit on $P_{\rm min}$, that is
independent of the EOS. This limit results from the condition of
causality, combined with the requirement that EOS yields neutron stars
with masses compatible with observed  ones [currently the highest
accurately measured neutron star mass is $M_{\rm obs}^{\rm
max}=1.442~{\rm M}_\odot$ (Taylor \& Weisberg 1989)].  It will be
hereafter referred to as $P_{\rm min}^{\rm CL}$.

The first calculation of $P_{\rm min}^{\rm CL}$ was done by Glendenning
(1992), who found the value of $0.33$~ms. Glendenning (1992), however,
used a rather imprecise empirical formula, to calculate a lowest
$P_{\rm min}$ by using the parameters (mass and radius) of the maximum
mass configurations of a  family of non-rotating neutron star models.
His result, therefore, should be considered only as an estimate of
$P_{\rm min}^{\rm CL}$. Recently, Koranda et al. (1997) extracted the
value of $P_{\rm min}^{\rm CL}$ from extensive {\it exact} calculations
of uniformly rotating neutron star models. They have shown, that the
method of Glendenning (1992) overestimated the value of $P_{\rm
min}^{\rm CL}$ by 6\%. The result of  Koranda et al. (1997)
calculations  can be summarized  in a formula
%%%%%%%%%
\begin{equation}
P_{\rm min}^{\rm CL} = 
0.196~{M_{\rm obs}^{\rm max}\over {\rm M}_\odot}~{\rm ms}~, 
\label{Koranda.Pmin}
\end{equation}
%%%%%%%%%%%%%%%
which combined with measured mass of PSR B1913+16 yields 
today's lower bound for $P_{\rm min}^{\rm CL}=0.282$ ms. 
This absolute bound on the 
minimum period was  obtained for  the  ``causality limit
(CL)  EOS''  
$p=(\rho-\rho_{\rm 0})c^2$, which yields neutron star models of the surface 
density $\rho_{\rm 0}$ and is maximally stiff (${\rm d}p/{\rm d}\rho = c^2$)  
everywhere within the star; it does not depend on the value of 
$\rho_0$. In the present letter we show that Eq. (1) 
can be reproduced  using an empirical  formula for $P_{\rm min}$ 
derived for 
{\it realistic} causal EOS by Lasota et al. (1996), 
combined  with an  upper bound on the relativistic 
(compactness) parameter  $2GM/Rc^2$ for static neutron stars with 
causal EOS. 
%
%%%%%%%%%%%%%%%%%%%%%%%%%%%%%%%%%%%%%%%%%%%%%%%%%%%%%%%%%
\section{Relation between $x_{\rm s}$ and $P_{\rm min}$}
As shown by Lasota et al. (1996), numerical results  of 
Salgado et al. (1994a,b) for the maximum frequency 
of uniform stable rotation can be reproduced (within better than 2\%),   
for a broad set of realistic causal EOS of dense 
matter, by an empirical formula
%%%%%%%%%%%%%%%%%%%%%%
\begin{equation}
\left(\Omega_{\rm max}\right)_{\rm e.f.}
 = {\cal C}(x_{\rm s})
\left(
{G M_{\rm s}\over R_{\rm s}^3}
\right)^{1\over 2}~,
\label{emp.form}
\end{equation}
%%%%%%%%%%%%%%%%%%%%%%%
\vskip 2mm
where $M_{\rm s}$ is the maximum mass of a spherical (nonrotating)
neutron star  
and $R_{\rm s}$ is the corresponding radius, and  
${\cal C}(x_{\rm s})$ is 
a universal (i.e. independent of the EOS) function  of the compactness 
parameter 
$x_{\rm s}\equiv 2GM_{\rm s}/R_{\rm s}c^2$ for the static maximum mass 
configuration, 
%%%%%%%%%%%%%%%%%%%%%%%%%%%
\begin{equation}
{\cal C}(x_{\rm s})= 
0.468 + 0.378 x_{\rm s}~. 
\label{C}
\end{equation}
%%%%%%%%%%%%%%%%%%%%%%%%%%%
Combining Eq. (2) and Eq. (3) we get
\vskip 2mm 
%%%%%%%%%%%%%%%%%%%%%%%%%%
\begin{equation}
\left(
P_{\rm min}
\right)_{\rm e.f.} 
= 
{8.754\times 10^{-2}\over {\cal C}(x_{\rm s})x_{\rm s}^{3\over 2}} 
~{M_{\rm s}\over {\rm M}_\odot}~{\rm ms}. 
\label{emp.rel}
\end{equation}
%%%%%%%%%%%%%%%%%%%%%%%%%%%%
\vskip 2mm
At given maximum mass of a spherical  configuration, the maximum  rotation 
frequency (minimum  rotation period) is obtained for the maximum 
value of $x_{\rm s}$.   At fixed $x_{\rm s}$, the value of $P_{\rm min}$ 
is proportional to $M_{\rm s}$. Neutron 
stars for which masses have been measured,  rotate so slowly
that their structure  
can be very well approximated by that of a spherical star. Observations impose 
thus a condition $M_{\rm s}\ge M^{\rm max}_{\rm obs}$. 
%%%%%%%%%%%%%%%%%%%%%%%%%%%%%%%%%%%%%%%%%%%%%%%%%%%%%%%%%
\section{Lower bound on $P_{\rm min}$   }
Our empirical relation, Eq. (4), indicates, that to minimize $P_{\rm
min}$ for given $M_{\rm obs}^{\rm max}$ we have to look for an EOS
which yields maximum $x_{\rm s}$ at $M_{\rm s}=M_{\rm obs}^{\rm max}$.
It is well known, that if one relaxes the condition of causality, the
absolute upper bound on $x_{\rm s}$ for stable neutron star models is
reached for an incompressible fluid  (i.e., $\rho=const.$) EOS; the
value of $x_{\rm s}$ is then independent of $M_{\rm s}$ and equal $8/9$
(see, e.g., Shapiro \& Teukolsky 1983).  It is therefore rather natural
to expect that in order   to maximize $x_{\rm s}$  under the condition
of causality, one has to maximize sound velocity throughout the star.
%####################### zmiany we fragmencie ponizej ###########
Together with condition of density continuity in the stellar interior
this  points out at the CL EOS, $p=(\rho-\rho_0)c^2$,  as to that which
yields ``maximally compact neutron stars''; 
introducing density
discontinuities does not increase the value of $x_{\rm s}$, see Gondek
\& Zdunik (1995). 
%#zmienilem poczatek nastepnego zdania - `remind' musi miec podmiot,
%ale tak czy inaczej lepiej jest bez niczego#%
[The conjecture that the CL EOS 
minimizes $P_{\rm min}$ was already proposed and then confirmed 
numerically in extensive exact calculations by Koranda et al. (1997)].
%############################ koniec zmienionego tekstu ######### 
Note, that the value of $x_{\rm s}$ for CL EOS does
not depend on $\rho_0$ (and therefore is $M_{\rm s}$-independent). It
represents an absolute upper bound on $x_s$ for causal EOS, $x_{\rm
s,max}$.  Our numerical calculation gives $x_{\rm s}({\rm
CL~EOS})=x_{\rm s,max}= 0.7081$. 
This corresponds to an absolute upper bound on the surface redshift 
of neutron star models with causal EOS, 
$z_{\rm max}=(1-x_{\rm s,max})^{-1/2} - 1 = 0.8509$. 

Let us consider the effect of the presence of a crust (more generally,
of an envelope of normal neutron star matter). For a given EOS of the
normal envelope, the relevant (small) parameter is the ratio $p_{\rm
b}/\rho_{\rm b}c^2$, where $p_{\rm b}$ and $\rho_{\rm b}$ are,
respectively, pressure and mass density at the bottom of the crust
(Lindblom 1984).  The case of $p_{\rm b}=0$ corresponds to stellar
models with no normal crust.  Numerical calculations show, that adding
a crust onto a CL EOS core implies an increase of $R_{\rm s}$, which is
linear in $p_{\rm b}/\rho_{\rm b}c^2 $; for a solid crust we have
typically $p_{\rm b}/\rho_{\rm b}c^2 \sim 10^{-2}$.  The change
(increase) in $M_{\rm s}$ is negligibly small; it turns out to be
quadratic in $p_{\rm b}/\rho_{\rm b}c^2$. This implies, that the
decrease of $x_{\rm s,max}$, and of the maximum surface redshift $z_{\rm
s,max}$, due to the presence of a crust,
is  proportional to $p_{\rm b}/\rho_{\rm b}c^2$. This   is consistent
with Table 1 of Lindblom (1984). However, the extrapolation of his
results to $p_{\rm b}=0$ yields $z_{\rm max}=0.891$, which is nearly
5\% higher than our value of $z_{\rm max}$ ! This might reflect a lack
of precision of the variational method used by Lindblom (1984), which
led to an overestimate of the value of $z_{\rm max}$.  It should be
stressed that while a precise determination of $M_{\rm max}\equiv
M_{\rm s}$ for static neutron star models is rather easy, determination
of the precise value value of the radius of the maximum mass
configuration, $R_{\rm s}$, (with the same relative precision as
$M_{\rm s}$) and consequently of the value of $x_{\rm s}$ (with, say,
four significant digits), is much more difficult and requires a rather
high precision of numerical integration of the TOV equations.

In what
follows, we restrict ourselves to the case of the absolute upper
bound on $x_{\rm s}$, obtained for neutron star models with no
crust.   
Inserting the value of $x_{\rm s,max}$ into  Eq. (4) we get 
%%%%%%%%%%%%%%%%%%%%%%%
\begin{equation}
\left(P^{\rm CL}_{\rm min}\right)_{\rm e.f.}
=0.1997 { M_{\rm obs}^{\rm max}\over {\rm M}_\odot}~{\rm ms}~.
\label{bound.emp}
\end{equation}
%%%%%%%%%%%%%%%%%%%%%%
Current lower bound  on $P$, resulting from the above equation, is thus 
$0.288$~ms, which is only 2\% higher than the 
result of extensive exact numerical 
calculations of Koranda et al. (1997).

The formula (\ref{bound.emp}) deserves an additional comment.  In
numerical calculations, of  a family of stable uniformly rotating
stellar models, for a given EOS of dense matter, one has to distinguish
between the rotating configuration of maximum mass, which corresponds
to the rotation frequency  $\Omega_{M_{\rm max}}({\rm EOS})$, and the
maximally rotating one, which rotates at $\Omega_{\rm max}({\rm EOS})$
(Cook et al. 1994, Stergioulas \& Friedman 1995). Notice, that
determination of a maximum mass rotating configuration (and therefore
of $\Omega_{M_{\rm max}}$)  is a much  simpler task than the
calculation of exact value of $\Omega_{\rm max}$, which is time
consuming and very demanding as far as the precision of numerical
calculations is concerned. Usually,  both configurations are very close
to each other, and $\Omega_{\rm max}$ is typically only 1-2\% higher
than  $\Omega_{M_{\rm max}}$; such a small difference is within the
typical precision of the empirical formulae for $\Omega_{\rm max}$.
Actually, the formula for ${\cal C}(x_{\rm s})$, Eq. (3),  was fitted
to the values of $\Omega_{M_{\rm max}}({\rm EOS})$ calculated in
(Salgado et al. 1994a,b). Therefore, Eq.(\ref{bound.emp}) should in
principle be used to evaluate the causal lower bound to $P_{{\rm
min},M_{\rm max}}$; it actually   reproduces, within 0.2\%,  the exact
formula for this quantity,   obtained by Koranda et al. (1997) [see
their Eq. (8)].

It should be stressed that Eq. (5) results from an {\it
extrapolation} of the  empirical formula of Lasota et al. (1996).
General experience shows that -  in contrast to interpolation -
extrapolation is a risky procedure. The fact  
that in our case extrapolation of an empirical formula
yields  - within 2\% -  the value of $P_{\rm min}$ 
 of Koranda et al. (1997) (and reproduces their value of 
$P_{{\rm min},M_{\rm max}}$), proves the usefulness of 
compact ``empirical expressions'' which might summarize, in a
quantitative way,   a relevant content  of extensive 
numerical calculations  of  uniformly rotating neutron star models.  
\begin{acknowledgements}
This research was partially supported  by the KBN grant No. 2P03D.014.13. 
 During his stay at DARC, P. Haensel was supported by the PAST Professorship  
of French MENESRT. 
\end{acknowledgements}
%%%%%%%%%%%%%%%%%%%%%%%%%%%%%%%%%%%%%%%%%%%%%%%%%%%%%%%%%%%
\par

\end{document}